\documentclass[rog]{agutex}

\usepackage{graphicx}
\usepackage{amsmath}
\usepackage{amsfonts}
\usepackage{amssymb}
\authorrunninghead{G. VERTH AND D. B. JESS}

\titlerunninghead{MHD wave modes resolved in the chromosphere}

\begin{document}

%
%

\title{MHD wave modes resolved in fine-scale chromospheric magnetic structures}
%
%

%
%



\authors{Gary Verth\altaffilmark{1} and David B. Jess\altaffilmark{2}}

\altaffiltext{1}{Solar Physics and Space Plasma Research Centre
(SP$^2$RC), The University of Sheffield, Hicks Building, Hounsfield
Road, Sheffield S3 7RH, UK}

\altaffiltext{2}{Astrophysics Research Centre,
School of Mathematics and Physics,
Queen's University Belfast, Belfast, Northern Ireland,
BT7 1NN, UK.}





%
%


\begin{abstract}
Within the last decade, due to significant improvements in the spatial and temporal resolution of chromospheric data, magnetohydrodynamic (MHD) wave studies in this fascinating region of the Sun's atmosphere have risen to the forefront of solar physics research.
In this Chapter we begin by reviewing the challenges and debates that have manifested in relation to MHD wave mode identification in fine-scale chromospheric magnetic structures, including spicules, fibrils and mottles. Next we go on to discuss how the process of accurately identifying MHD wave modes also has a crucial role to play in estimating their wave energy flux.  This is of cardinal importance for estimating what the possible contribution of MHD waves is to solar atmospheric heating.  Finally, we detail how such advances in chromospheric MHD wave studies have also allowed us, for the first time, to implement cutting-edge magnetoseismological techniques that provide new insight into the sub-resolution plasma structuring of the lower solar atmosphere.
\end{abstract}

%
%

%

\begin{article}

%
%

\section{Introduction}
\label{intro}
Due to its complex and dynamic fine-scale structure, the chromosphere is a particularly challenging region of the Sun's atmosphere to
understand \citep[see, e.g.,][]{Jud06}. It is now widely accepted that to model chromospheric dynamics, even on a magnetohydrodynamic (MHD) scale, while also calculating spectral line emission, one must realistically include the effects of partial ionization and radiative transfer in a multi-fluid plasma under non-LTE conditions \citep[e.g.,][]{Han07}. Within the past decade there has been a concerted international effort to try and advance our understanding of this tantalising layer of the solar atmosphere, which is thought to be a key part of solving the solar atmospheric heating problem. There have certainly been major advances in chromospheric observations from high spatial and temporal resolution space-borne and ground-based instruments. These include Hinode \citep{Kos07} launched in 2006 and the Rapid Oscillations in the Solar Atmosphere \citep[ROSA;][]{Jes10ROSA}
multi-wavelength camera system based at the Dunn Solar Telescope (DST), which became operational in 2009. Furthermore, in 2013 the Interface Region Imaging Spectrograph \citep[IRIS;][]{DeP14} was launched with the specific task of studying the previously little explored UV lines formed in the interface region between the chromosphere and corona in unprecedented detail.

There now seems to be general agreement that the Sun's magnetic field is primarily responsible for plasma heating in its atmosphere, from the photosphere, up through the chromosphere, interface region and finally into the corona. However, a rigorous debate is still ongoing as to which particular plasma processes are actually responsible.  Historically, popular proposed mechanisms such as MHD wave dissipation or magnetic reconnection have had little direct observational evidence to support them due to the small spatial scales involved and the limited resolution of past instrumentation. For example, earlier space-borne EUV and X-ray wavelength instruments launched in the 1990's, with their limited spatial and/or temporal resolutions, e.g., the SOlar and Heliospheric Observatory \citep[SOHO;][]{Dom95} and the Transition Region and Coronal Explorer \citep[TRACE;][]{Han99}, could only detect less frequent high energy reconnection/wave events. However, it is now apparent that the majority of heating in the solar atmosphere must be taking place on small spatial scales, certainly less than $\sim 100$ km perpendicular to the magnetic field direction. As a result, the rarity of large-scale reconnection/wave events related to, e.g., flares and CMEs means that these do not play a key role.

Regarding the chromosphere, since it can be observed from ground-based telescopes with much higher spatial and temporal resolutions than are currently possible for the corona, it has presented an opportunity to study the small-scale and ever-present dynamics of thin ($ \lesssim 1000$~km wide) chromospheric magnetic features such as spicules, fibrils and mottles. These are particularly visible in narrowband spectral filters such as H$\alpha$ and Ca {\sc{ii}} H and K, which have aided tremendously in the identification of various MHD wave modes propagating along such fine-scale chromospheric magnetic structures. Prior to these modern high-resolution chromospheric observations, an abundance of MHD waves were expected to exist in the Sun's lower atmosphere since it is in essence an elastic/compressible medium permeated by strong magnetic fields that are constantly being stressed and perturbed by the magneto-convective motions generated below (see Chapter~25 for an overview of photospheric wave modes). In agreement with these predictions it has now been confirmed that there are indeed both Alfv\'{e}nic and magneto-acoustic wave modes propagating along chromospheric waveguides at all times \citep[e.g.,][to name but a few]{DeP07,He09a, He09b, Mor12a,Kur12,Kur13}.

A significant breakthrough now is that with contemporary multi-instrumental studies we can see how small-scale disturbances generated in the photosphere impact on the higher atmospheric layers. For example, in the particular case of ubiquitous small-scale ($\approx$ 1000 km diameter) photospheric vortical motions, \citet{Wed12} traced the resultant energy transfer of associated magnetic tornadoes up through the atmosphere by exploiting simultaneous photospheric, chromospheric and coronal observations using the ground-based CRisp Imaging Spectropolarimeter \citep[CRISP;][]{Sch08} at the Swedish Solar Telescope (SST) and also the Atmospheric Imaging Assembly \citep[AIA;][]{Lem12} on board the Solar Dynamics Observatory (SDO). SDO, launched in 2010, is especially useful in detecting the coronal signatures of energy propagation from the lower atmosphere since it continually observes the full solar disc, sampling UV/EUV emission about every 10~seconds. This now makes it much easier for observers to compare co-spatial/temporal coronal data to photospheric/chromospheric data gathered from specific, limited duration ground-based campaigns.

With all these new possibilities, there has been a step-change in all-encompassing solar atmospheric MHD wave studies in the past decade. Now in simultaneous multi-wavelength observations it is  possible to observe abundant chromospheric waves, along with their photospheric drivers and associated
coronal signatures \citep[e.g.,][]{Mor14}. However, before we can start considering the energetics of the waves and their possible contribution to plasma heating, the very first step that must be taken is to accurately identify which wave modes are being observed. This is not a trivial task and has actually been the cause of much debate since the first wave interpretations of Hinode's chromospheric data \citep[see, e.g.,][]{Erd07,Van08}. However, without this knowledge it not possible to quantify what the wave cut-off behaviour and most likely damping mechanisms could be. In fact, specific MHD wave modes can have different dominant (or competing) frequency-dependent damping mechanisms, e.g., thermal conduction, radiative cooling, kinematic viscosity, MHD radiation, resonant absorption and phase mixing \citep[see, e.g.,][]{Asc04}. This can result in wildly varying damping rates for different MHD wave modes and, in turn, their ultimate effectiveness for atmospheric heating. Hence, accurate quantification of MHD wave energetics must be founded on an precise identification of the actual wave mode (or combination of wave modes) being observed, as documented in the recent review by \citet{Jess15}.

\begin{figure*}
\begin{center}
\includegraphics[angle=0,width=14cm]{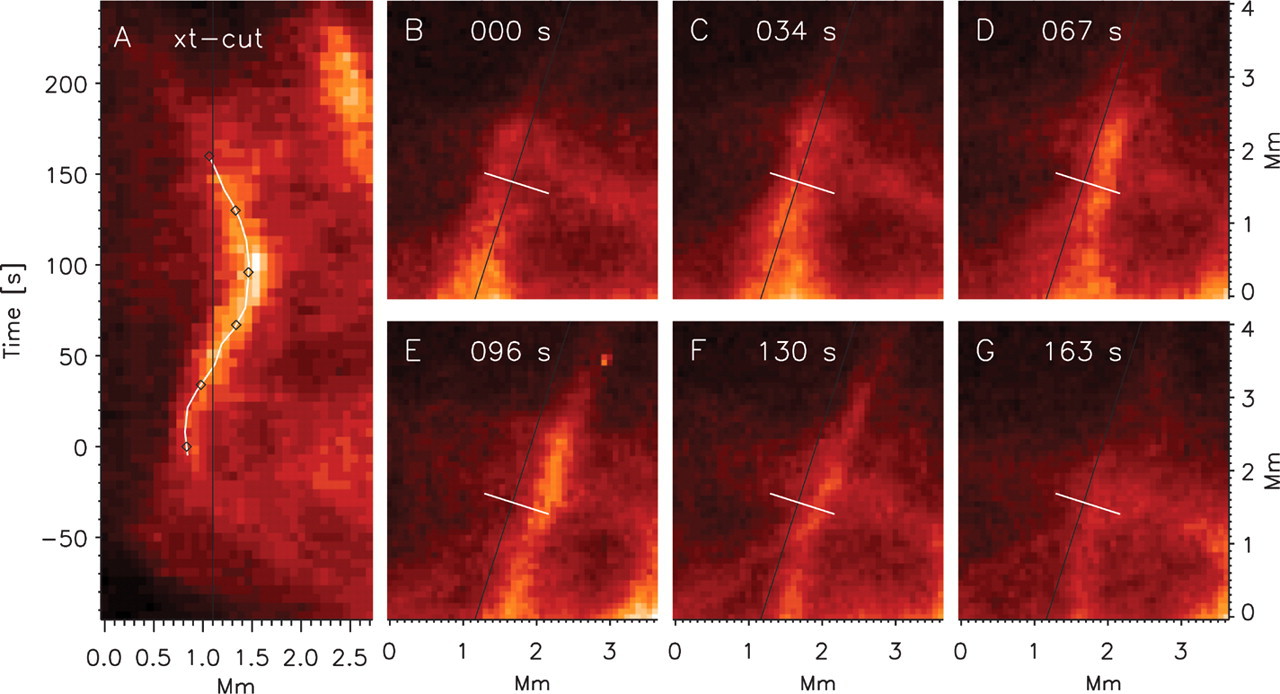}
\end{center}
\caption{Example of an MHD kink wave observed off-limb in a spicule using the Ca~{\sc{ii}}~H filter of Hinode/SOT by \citet{DeP07}. The panels demonstrate the plane-of-sky projection of the spicule's motion. The largest (left-hand) panel shows a time--distance diagram taken at a slice approximately perpendicular to the spicule axis. The smaller panels reveal sequential snapshots of the spicule and indicate the perpendicular slice as a white line. Image reproduced from \citet{DeP07}.}
\label{kink_spicule}
\end{figure*}

Most recently, the benefits of accurate wave mode identification have been threefold; first they allow us to more precisely quantify the chromospheric energy flux associated with each mode \citep{Mor12a, Van14}; second, the broadband frequency information of  MHD waves allows us to study the frequency-dependent wave damping \citep{Ver10, Mor14}; thirdly, it allows a complimentary approach to understanding the fine-scale plasma structure of the chromosphere by implementing magnetoseismological techniques \citep{Fed11b,Ver11, Mor12b,Kur13,Mor14a}. These are all crucial gains that were only made possible in the last decade due to the launch of Hinode, combined with significant improvements in ground-based spectroscopic/imaging observations (ROSA/DST and CRISP/SST) and polarimetry, most notably with the Coronal Multi-channel Polarimeter \citep[CoMP;][]{Tom09}. Now, with the launch of
SDO and IRIS, there is an urgent need to combine the best cutting-edge observational and modelling studies so that forward leaps in solar physics can be facilitated. This will open a whole new era of studying the heat generated through fine-scale plasma dynamics in the solar atmosphere, not just through the study of waves, but also, e.g., instabilities and nanoflare heating events. The following Sections~\ref{kink} to \ref{torsional} deal with MHD wave mode identification in chromospheric waveguides. Then, in Section~\ref{energy flux} we go on to consider energy flux estimates of particular wave modes. Finally, in Section~\ref{CMS} we also detail how contemporary chromosopheric wave observations have been exploited for the purposes of advancing the field of solar atmospheric magnetoseismology. We would like to note that in solar physics literature the term ``magnetoseismology'' is also known as ``MHD seismology'', or just simply ``seismology''. However, for the purposes of consistency we will employ the term ``magnetoseismology'' throughout this Chapter.

\section{MHD kink mode identification}
\label{kink}

Although quasi-periodic line widths and Doppler velocities were detected in solar spicule data as far back as the 1960s \citep[e.g.,][]{Nik67, Pas68, Wea70}, the limited spatial and temporal resolutions of this era prevented the identification of specific MHD wave modes. Spicules have actually been of much interest to solar physicists for much longer. In fact they were first reported in scientific literature as far back as \citet{Sec77}. Nowadays they are known to be thin jets of plasma channelled by the magnetic field in the Sun's lower atmosphere \citep[see, e.g., the review by][]{Zaq09}. However, what causes their formation and what their contribution is to plasma heating (if any) is still the matter of fierce debate \citep[see, e.g.,][]{DeP11, Kli12}.

Spicules, which are predominantly rooted at network boundaries, are seen off-limb in a 2D projection as a ``thick forest'' in chromospheric lines such as H$\alpha$ and Ca~{\sc{ii}}~H and K. The first claim of kink wave detection in spicules was by \citet{Kuk06} using H$\alpha$ Doppler data from the coronagraph and universal spectrograph based at the Abastumani Astrophysical Observatory in Georgia. However, it took high resolution imaging data from the Solar Optical Telescope \citep[SOT;][]{Sue08, Tsu08b} onboard Hinode to observe this particular MHD wave mode more unambiguously \citep{DeP07}. Periodic motions, perpendicular to the direction of the magnetic field, were detected in spicules using the Ca~{\sc{ii}}~H filter of SOT (see Figure~\ref{kink_spicule}). This strongly suggested that the main restoring force for these waves was magnetic tension and led \citet{DeP07}, and later \citep{He09a}, to simply interpret them as Alfv\'{e}n waves \citep{Alf42} with their phase speed, $c_A$, governed by the well known relation,
\begin{equation}
c_A=\frac{B}{\sqrt{\mu \rho}} \ ,
\end{equation}
 where $B$ is the magnetic field strength, $\mu$ is the magnetic permeability and $\rho$ is the plasma density. Theorists immediately started debating the validity of this interpretation \citep[see, e.g.,][]{Erd07,Van08}. The base objection was that Alfv\'{e}n's linear wave and planar geometry model assumed the plasma to be completely homogeneous, and was therefore not accurate enough to predict the observed properties of waves travelling through the Sun's inhomogeneous and finely structured atmosphere. In fact, spicules have a finite width (diameter $ \lesssim 1000$ km) and likely have a substantial variation in plasma density transverse to the direction of the magnetic field \citep[e.g.,][]{Bec68}, further fuelling the debate as to whether bulk Alfv{\'{e}}n waves were the correct interpretation.

Recent 3D MHD radiative transfer simulations suggest that spicules could be formed by a localised and enhanced Lorentz force at their base, which squeezes the chromospheric plasma in such a way that it is thrown up to lower coronal heights \citep{Mar11, Mar13}.  Such cutting-edge numerical modelling supports the idea of spicules being overdense relative to the ambient plasma. If one assumes that on average spicules represent thin magnetically dominated filaments of plasma, with chromospheric densities and temperatures that penetrate into the corona, then the observed transverse motions indeed have to be modelled as MHD waves propagating along an overdense flux tube relative to the ambient plasma. Models of this type in flux tube, i.e., cylindrical, geometry have been around since the late 1970s \citep[e.g.,][]{Zai75, Wen79, Wil79, Wil80, Edw83}. Deriving the dispersion relations in such models requires the physical constraints that the total pressure perturbations and normal velocity components be continuous at the flux tube boundary. This was a worthwhile advance on Alfv\'{e}n's simple model since it allowed for more realistic geometry and the possibility of both transverse magnetic field and plasma density inhomogeneities. The derived dispersion relations resulted in a much richer variety of MHD wave modes than was possible in Alfv\'{e}n's more simple model. Assuming that the equilibrium plasma variation in the azimuthal direction of the flux tube is negligible, it allows for Fourier decomposition in that direction, and ultimately wave mode categorisation in terms of the integer azimuthal wave number, $m$. For a cylindrical flux tube, the lowest order azimuthal wavenumber ($m=0$) results in two distinct decoupled MHD axisymmetric wave modes, i.e., the incompressible torsional Alfv\'{e}n \citep[see, e.g.,][]{Hol78} and the compressible sausage \citep[see, e.g.,][]{Nak03, Asc04a} modes. Their specific defining physical properties and their recent identification in chromospheric observations are discussed in Sections~\ref{sausage} \& \ref{torsional}.

\begin{figure*}
\begin{center}
\includegraphics[angle=0,width=11.5cm]{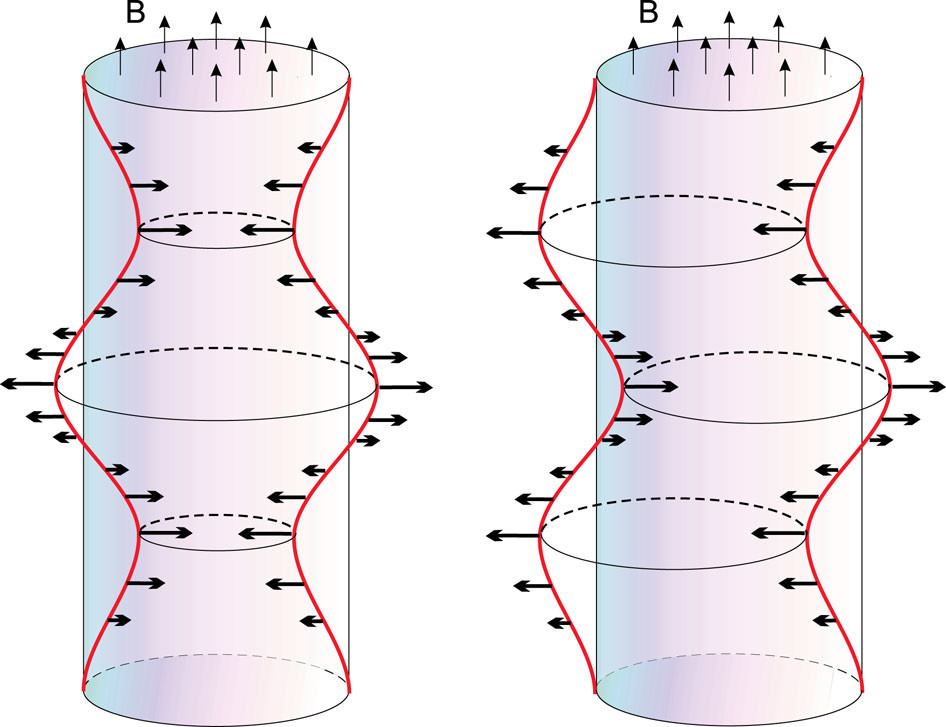}
\end{center}
\caption{Cartoon illustrating the MHD sausage and kink waves in a magnetic flux tube. The sausage wave (left), which has $m=0$, is characterized by an axi-symmetric contraction and expansion of the tube's cross-section. This produces a periodic compression/rarefaction of both the plasma and magnetic field. The kink wave (right) has an azimuthal wavenumber $m=1$. This is notable since it is the only value of $m$ that causes a transverse displacement of the flux tube. In contrast to the sausage wave, the kink wave displacement/velocity field is not axi-symmetric about the flux tube axis. The red lines show the perturbed flux tube boundary and thick arrows show the corresponding displacement vectors. The thin arrows labelled B show the direction of the background magnetic field. Image from \citet{Mor12a}.}
\label{saus_kink}
\end{figure*}

After $m=0$, the next integer azimuthal wavenumber is $m=1$ and this is associated with the so-called kink mode, which is the particular MHD wave mode under discussion in this Section. As shown in Figure~\ref{saus_kink} (right), the key feature of this $m=1$ mode is that it is the only value of $m$ that produces a bulk transverse displacement of the flux tube. The $m=0$ and all higher order ($m\ge 2$) fluting modes do not do this.
Because of the plasma structuring the kink speed depends on both the internal magnetic field strength ($B_i$) and plasma density ($\rho_i$), as well as the external magnetic field strength ($B_e$) and plasma density ($\rho_e$). In the zero plasma-$\beta$ limit equilibrium demands $B_e=B_i$. In this magnetically dominated plasma regime the kink speed, $c_k$, is described in the thin tube (or long wavelength) limit as,
 \begin{equation}
 c_k=B\sqrt{\frac{2}{\mu(\rho_i+\rho_e)}} \ ,
 \end{equation}
where $B=B_e=B_i$. Note that the value of the kink speed lies between that of the internal and external Alfv\'{e}n speeds. The kink mode is highly Alfv\'{e}nic since its main restoring force is magnetic tension \citep[see the discussion by][]{Goo09}.  As discussed by \citet{Van08}, it is also only weakly compressible in the long wavelength regime, hence it is unlikely that intensity perturbations due to compression/rarefaction could be readily observed for such waves. Actually, \citet{DeP07} did not report any intensity perturbations concurrent with the transverse waves observed in spicules. Therefore, if present, they must be very small relative to the background  intensity. So it is worthwhile to note that the absence of detectable intensity perturbations cannot be used as an argument to discount the presence of kink waves in favour of bulk Alfv\'{e}n waves.
\citet{Erd07} and \citet{Van08} also criticised the Alfv\'{e}n wave interpretation of \citet{DeP07} from the flux tube perspective. Since Alfv\'{e}n waves must be torsional in flux tube geometry, they would not display the bulk transverse motions observed in the imaging data of Figure~\ref{kink_spicule}, i.e., the flux tube would appear stationary. This was the main argument put forward to support the kink mode interpretation over the initial Alfv\'{e}n wave interpretation suggested by \citet{DeP07}.

Regarding observed properties of kink waves in spicules, in a case study of 94 events, \citet{DeP07} reported periods of $100-500$~s and transverse velocity amplitudes on the order of $10-25$~km{\,}s$^{-1}$. A subsequent, but more limited case study by \citet{He09a}, also with Hinode/SOT data, reported the presence of higher frequency kink waves with periods as short as $40-50$~s but with similar velocity amplitudes to that estimated by \citet{DeP07}. An advance on the work of \citet{DeP07} by \citet{He09a} was an attempt to actually measure the upward propagation speed of the waves, resulting in values between $60-150$~km{\,}s$^{-1}$ up to 7~Mm above the solar limb. A larger case study of 89 spicules by \citet{Oka11} found a median period of 45~s, which is more consistent with the period estimates of \citet{He09a} than \citet{DeP07}. Interestingly, \citet{Oka11} also reported that 59\% of the waves were propagating upwards, 21\% propagating downwards and 20\% standing. \citet{Oka11} also estimated propagation speeds of $164-267$~km{\,}s$^{-1}$ up to 7 Mm above the limb, but stated that inferred speeds greater than 1000 km{\,}s$^{-1}$ above 10 Mm could not be physical. A possible explanation of this is that spicule intensity in chromospheric lines is too diffuse at higher altitudes to be reliable enough for wave studies.

Concerning another important class of chromospheric fine-scale magnetic structures, statistical studies of kink waves in fibrils have been made in a series of papers by \citet{Mor12a, Mor13, Mor14}. Fibrils are low-lying elongated structures, most clearly seen on disc, that span supergranular cells \citep{Fou71,Zir72}. Like spicules, they too have narrow widths ($\lesssim$~1000 km) and therefore need the highest resolution instruments available to analyse their wave properties. To this end, a range of statistical studies were performed by \citet{Mor12a} to exploit the fantastic capabilities of ROSA \citep{Jes10ROSA} equipped with a narrowband (0.25~{\AA}) H$\alpha$ filter. \citet{Mor12a} found that bulk transverse oscillations in fibrils, as in spicules, were omnipresent. It was reasoned that because fibrils appear dark in the line core of H$\alpha$, it was most likely a result of a density enhancement relative to the ambient plasma that causes the radiative emission from the photosphere to suffer increased dimming in their vicinity \citep{Pie11}. Hence, \citet{Mor12a} classed fibrils as overdense waveguides and that their bulk transverse oscillations must, along with spicules, be interpreted as kink waves, not Alfv\'{e}n waves. The combined studies of \citet{Mor12a, Mor13, Mor14} analysed 1688 fibril kink wave events in both the quiet Sun and active region chromospheres. The collated results gave most kink wave periods in the range of $94-130$~s with velocity amplitudes of $5-25$~km{\,}s$^{-1}$, which are certainly of the same order as that found in the aforementioned observations of kink waves in spicules. Perhaps this should be expected since both structures are rooted in intergranular lanes where it is likely they are excited by similar/identical drivers.

Kink waves on-disc have also been detected in mottles and via Rapid Blue-shifted Excursions observed in the blue wings of chromospheric spectral lines. Apart from their wave properties, mottles and Rapid Blue-shifted Excursions are of particular interest to solar physicists since it is thought they may be related directly (or indirectly) to spicules.  Mottels, like spicules and fibrils, are also thin magnetically aligned structures less than 1000~km wide, but can either appear dark or bright in chromospheric lines. They were identified in scientific literature as far back as the early 1970s \citep[e.g.,][]{Ali71, Saw72}. ROSA investigations of kink waves in mottles by \citet{Kur12, Kur13} revealed transverse velocity amplitudes on the order of $8-11$~km{\,}s$^{-1}$, which again are of the same order as found in spicules. The same is also true of kink waves detected in Rapid Blue-shifted Excursions in large scale statistical studies by \citet{Rou09} and \citet{Sek12,Sek13}, where they found transverse velocity amplitudes in a similar range. The observational characteristics of kink waves in all the various chromospheric waveguides discussed in this Section are summarised in Table~\ref{kink_tab}.

\begin{table*}[t]
\label{kink_tab}
\tiny
\begin{center}
\caption{Average (or individually measured) properties of chromospheric MHD kink waves.}
\begin{tabular}{llcccccl}
\hline
& & & & & & & \\
Structure & Region & {Max. displacement} & {Period} & {Max. velocity } & {Kink speed} & No. Events & {Reference} \\ [0.3ex]

& & {amplitude (km)} & {(s)} & {amplitude (km{\,}s$^{-1}$)} & {(km{\,}s$^{-1}$)} & &  \\ [0.3ex]
\hline \\ [0.3ex]
Spicule & CH &  $200-500$ & $150-350$ & $20\pm 5$ & -   & 95 &   \citet{DeP07} \\
 &CH &  -  & $60-240$ & $20\pm 5$ & -    &  -&   \citet{Sue08} \\
 & CH &  $1000$ & $130$ & $15$ & 460    &  1 &   \citet{Kim08} \\
&&  $700$ & $180$ & $8$  & 310   &   1 &     \\
&&  $800$ & $170$ & $9$  & 260   &   1 &     \\
 & CH &  $36$ & $48$ & $4.7$ & 75-150  &  1 &     \citet{He09a} \\
 &&  $36$ & $37$ & $6.1$ & 59-117  &   1 &    \\
 &&  $130$ & $45$ & $18.1$ & 73   &   1 &     \\
 &&  $166$ & $50$ & $20.8$ & 109-145  &  1 &    \\
& CH &  $55\pm 50$ & $45\pm 30$ & $7.4\pm 3.7$ & 160-305 &  89 &     \citet{Oka11} \\
& &  $600$ & $180$ & $22$ & - &  1 &     \citet{Eba12} \\
 & QS &  $670$ & $220$ & $19.2$ & - &  1 &    \citet{Jes12}\\
& &  $630$ & $139$ & $28.3$ & - &  1 &     \\
& &  $160$ & $65$ & $14.8$ & - &  1 &      \\
& &  $410$ & $158$ & $16.2$ & - &  1 &      \\
& &  $380$ & $129$ & $18.5$ & - &  1 &      \\
&&  $200$ & $105$ & $11.8$ & - &  1 &      \\
& &  $190$ & $171$ & $7.2$ & - &  1 &      \\
&AR&  $283\pm 218$ & $ - $ & $14\pm 112$ & - &  112 &    Type-I -\citet{Per12}\\
&AR&  $463\pm 402$ & $ - $ & $18\pm 12$ & - &  58 &    Type-II\\
&QS&  $245\pm211$ & $-$ & $16\pm 11$ & - &  174 &    \\
&CH&  $342\pm257$ & $-$ & $20\pm 12$ & - &  170 &    \\
\hline \\ [0.3ex]
Fibrils  &&  $135$ & $135$ & $1$ & 190  & 1 & \citet{Pie11}   \\
 & QS &  $315\pm 130$ & $-$ & $6.4\pm 2.8$ & 50-90   & 103  & \citet{Mor12a}   \\
 &QS &  $71\pm 37$ & $94\pm 61$ & $24.5\pm 1.8$ & -   &  & \citet{Mor13}   \\
& QS &  $94\pm 47$ & $116\pm 59$ & $5.5\pm 2.4$ & -  & 841  & \citet{Mor14}   \\
& AR &  $73\pm 36$ & $130\pm 92$ & $4.4\pm 2.4$ & - & 744   &   \\
\hline \\ [0.3ex]
Rapid Blue-shifted  & &  $300$ & $-$ & $8$ & -   &  35&   \citet{Rou09} \\
Excursions & CH &  $200$ & $-$ & $4-5$ & -   &  960&   \citet{Sek12} \\
 & QS &  $200$ & $-$ & $8.5$ & -   &  1951 &  average -  \citet{Sek13} \\
 & &  $220$ & $-$ & $11.7$ & -   &  1951 &  maximum  \\
\hline \\ [0.3ex]
Mottles & QS & $200\pm 67$  & $165\pm 51$ & $8.0\pm 3.6$ & -   &  42&   \citet{Kur12} \\
 & QS &  $\sim 172$ & $120\pm 10$ & $\sim 9$ & 50   &  1&   \citet{Kur13} \\
 & QS &  $252$ & $180\pm 10$ & $8.8\pm 31$ & $101\pm 14$  &  1&    \\
 & QS &  $327$ & $180\pm 10$ & $11.4\pm 3.3$ & $79\pm 8$   &  1&    \\
\hline
\end{tabular}
\end{center}
\end{table*}

\section{MHD Sausage mode identification}
\label{sausage}
\begin{figure*}
\begin{center}
\includegraphics[angle=0,width=12cm]{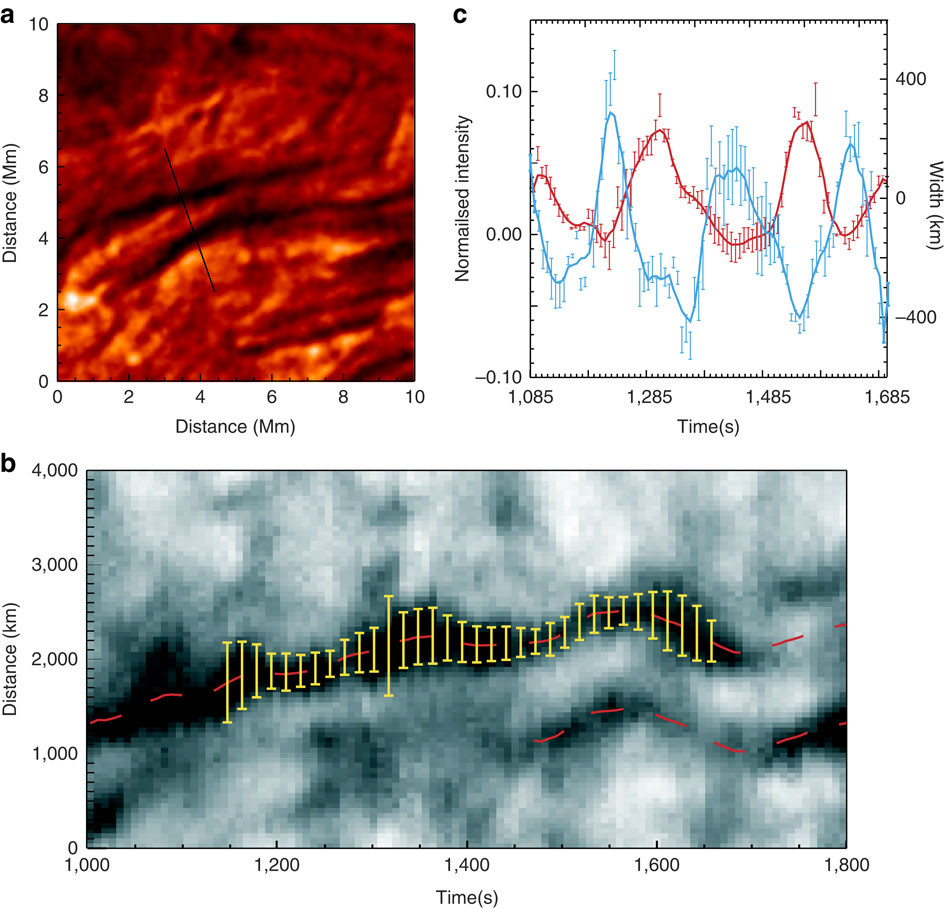}
\end{center}
\caption{Panel (a) depicts a cropped ROSA H$\alpha$ snapshot containing a pair of dark, and hence dense, chromospheric flux tubes. Using the cross-cut (black line) to extract intensity information, panel (b) displays the resulting time--distance diagram revealing the dynamic motion of the waveguides. Times are given in seconds from the start of the data set, while the overplots are the results from a Gaussian fitting routine to show concurrent kink (red line shows the central axis of the structure) and sausage waves (yellow bars show the measured width of structure). Here there are counter-propagating kink waves with periods of $232 \pm 8$ s and phase speeds of $71 \pm 22$~km{\,}s$^{-1}$ upwards and $87 \pm 26$~km{\,}s$^{-1}$ downwards. The maximum transverse velocity amplitudes in both cases is about  $5$~km{\,}s$^{-1}$. The sausage wave shown here has a period of $197 \pm 8$~s, a phase speed of $67 \pm 15$~km{\,}s$^{-1}$ and maximum transverse velocity amplitude of $1 - 2$~km{\,}s$^{-1}$. Panel (c) displays a comparison between the detected intensity (blue) and width (red) perturbations resulting from the Gaussian fitting. Sausage waves can naturally cause such anti-phase behaviour. Image reproduced from \citet{Mor12a}.}
\label{ROSA_saus_kink}
\end{figure*}

Unlike the weakly compressible non-axisymmetric kink mode, the axi-symmetric sausage mode is highly compressible, producing periodic changes to the cross-sectional area of a magnetic flux tube, analogous to fluid motion driven in an elastic tube by a peristaltic pump, as shown in Figure~\ref{saus_kink} (left). The presence of such motion was first detected in the Sun's lower atmosphere at the photospheric level in solar pores by \citet{Dorotovic2008} employing the G-band filter of the SST, and subsequently by \citet{Fuj09} using Hinode/SOT. These intense magnetic features are essentially like small sunspots without penumbrae. Employing ROSA G-band data, \citet{Mor11}  accurately measured the periodic area and intensity changes exhibited by solar pores. It was found that some pores exhibited a clear anti-phase behaviour between area and intensity oscillations that was strongly indicative of the sausage mode. For further and more in-depth discussions of photospheric sausage mode observations see Chapter 25. In the context of the current Chapter, these initial photospheric discoveries naturally led to the search for sausage modes higher up in the chromosphere. Using ROSA H$\alpha$ data, \citet{Mor12a} did indeed detect the anti-phase behaviour of flux tube width changes and intensities in fibrils (see Figure~\ref{ROSA_saus_kink}). Furthermore, these sausage waves were found to be concurrent with kink waves (whose identification in fibrils was previously discussed in Section~\ref{kink}). \citet{Mor12a} found the sausage waves to have periods in the range $135-241$~s and transversal velocities on the order of $1-2$~km{\,}s$^{-1}$. The resulting comparisons that can be made between kink/sausage energy fluxes in fibrils will be discussed in Section~\ref{energy flux}.

Slightly earlier work by \citet{Jes12}, again with ROSA H$\alpha$ data, actually detected similar joint kink and sausage mode signatures in spicules seen on-disc. However, \citet{Jes12} did not explicitly associate the observed patterns as being concurrent sausage/kink waves. The authors were primarily concerned with the footpoint driving mechanisms of these chromospheric waves since their dataset had simultaneous photospheric (G-band) and lower-chromospheric (Ca~{\sc{ii}}~K) image sequences. G-band photospheric intensity oscillations showed a distinct phase difference of about $90^{\circ}$ across a magnetic bright point located at the footpoint of a group of chromospheric spicules. A 2D MHD simulation was performed by \citet{Jes12} using a compressive field-aligned footpoint driver of maximum amplitude 12.5~km{\,}s$^{-1}$ combined with the observed spatial phase difference. This actually resulted in combined kink and sausage waves similar to those observed in spicules by \citet{Jes12} and fibrils by \citet{Mor12a}. Hence, this could provide an explanation of why both highly and weakly compressive wave modes occur together in such fine-scale chromospheric structures. Thus far, sausage waves have not been detected in off-limb spicules. This could be due to complicated line-of-sight effects present in the ``thick forest'' of spicules seen off-limb. However, for better understanding of spicule wave dynamics and energetics, proving their existence (or not) should certainly be the focus of future studies.

\begin{figure*}
\begin{center}
\includegraphics[angle=0,width=12cm]{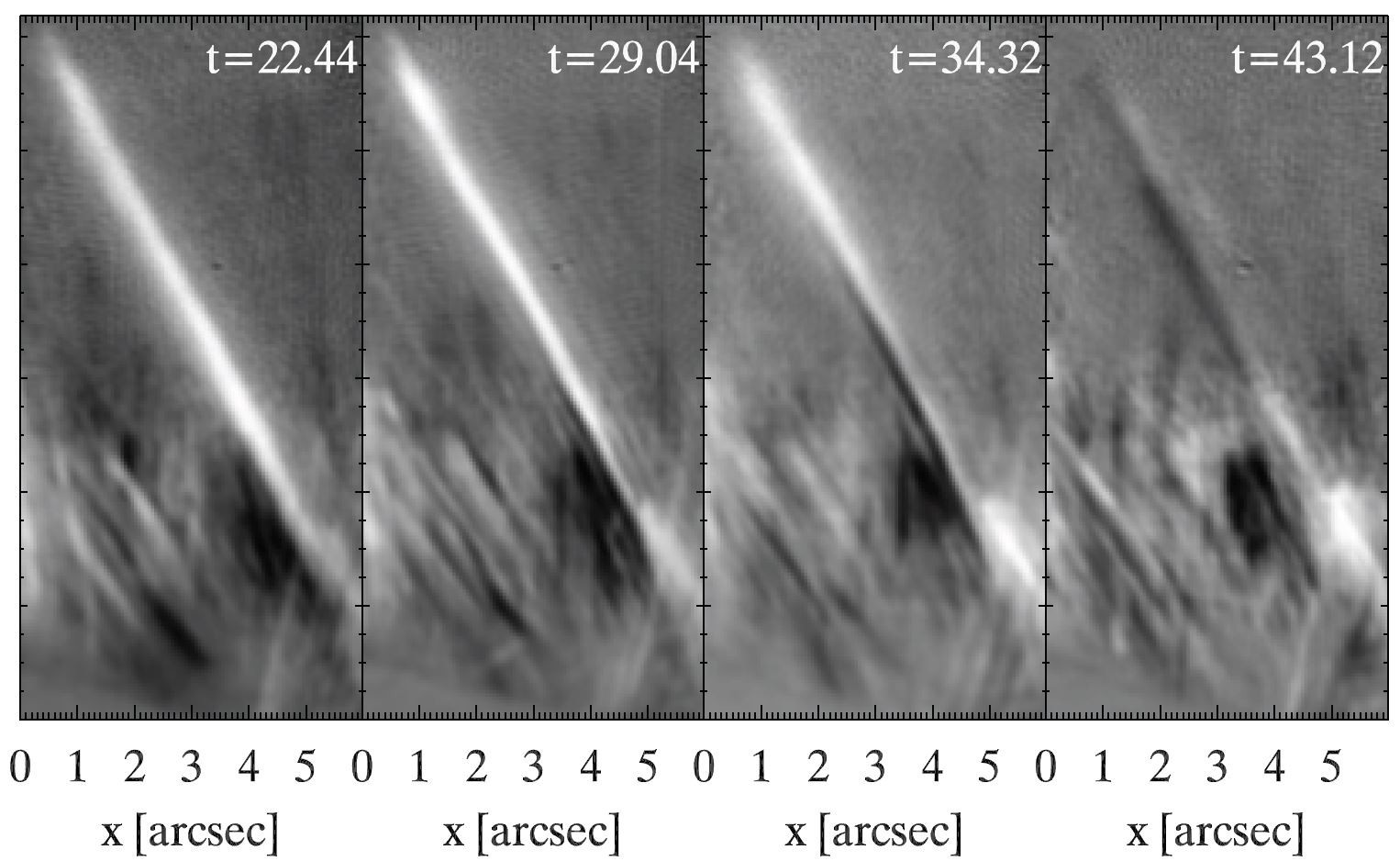}
\end{center}
\caption{Panels contain sequences of four CRISP H$\alpha$ Doppler images of a spicule at different sampling times (in seconds, blueshift bright). The largest Doppler shifts in the spicule are approximately $\pm 55$~km~s$^{-1}$. Image adapted from \citet{DeP12}.}
\label{torsion_doppler}
\end{figure*}

\section{MHD torsional Alfv\'{e}n wave \\ identification}
\label{torsional}
Observations of small-scale photospheric vortical motions, detected via G-band bright point tracking in intergranular lanes, have been the subject of much interest in recent years \citep[e.g.,][]{Bon08, Bon10, Wed09, Stei10, Wed12, Mor13}. G-band bright points, of the order 200~km in diameter, are often co-spatial with kG magnetic flux concentrations. In such cases they are often referred to as magnetic bright points \citep[e.g.,][]{Ste85, Sol93, Cro09, Jes10, Key11}.
Such small scale intense magnetic flux tubes rooted in vortex flow fields are natural sources of torsional Alfv\'{e}n waves, as well as other MHD wave modes \citep[e.g.,][]{Fed11a,She13}.

Torsional Alfv\'{e}n waves, if propagating in a near or sub-resolution flux tube could cause identifiable simultaneous periodic red and blue shifts in an observed spectral line. Since this process is not necessarily related to wave/energy dissipation inducing temperature fluctuations, it falls under the guise of periodic non-thermal spectral line broadening, providing the torsional amplitudes are large enough to cause noticeable red/blue shifts \citep{Zaq09}.
With H$\alpha$ data from the Solar Optical Universal Polarimeter (SOUP) based at the SST, \citet{Jes09} detected such periodic spectral line broadening above a magnetic bright point group. Since there was an absence of both co-spatial intensity oscillations and bulk transverse motions, the periodic spectral line broadening was interpreted by \citet{Jes09} as evidence of torsional Alfv\'{e}n waves. The estimated periods were in the range $126-700$~s and the average line-of-sight velocity amplitude was approximately 1.5~km{\,}s$^{-1}$.

Theoretically, torsional Alfv\'{e}n waves can exist for any azimuthal wavenumber, $m$, and by definition they are completely incompressible. A question that arises from interpreting observations as torsional Alfv\'{e}n waves is: {\it{How likely is it that such purely divergence-free MHD wave modes are actually excited in the photosphere?}} Follow up 3D MHD numerical investigations by \citet{Fed11b} of flux tubes driven by vortex drivers demonstrated that although torsional Alfv\'{e}n waves could be the dominant wave mode, kink and sausage waves were still unavoidably present. This was due to the fact that the chosen spiral driver did not have the particular azimuthal symmetry (or asymmetry) of one distinct $m$ value, but was in fact a superposition of different $m$ values. Feature tracking studies to estimate horizontal velocity field components at vortex locations in the photosphere \citep[e.g.,][]{Mor13} show that this complicated scenario is actually much closer to reality.

\begin{figure*}
\begin{center}
\includegraphics[angle=0,width=16cm]{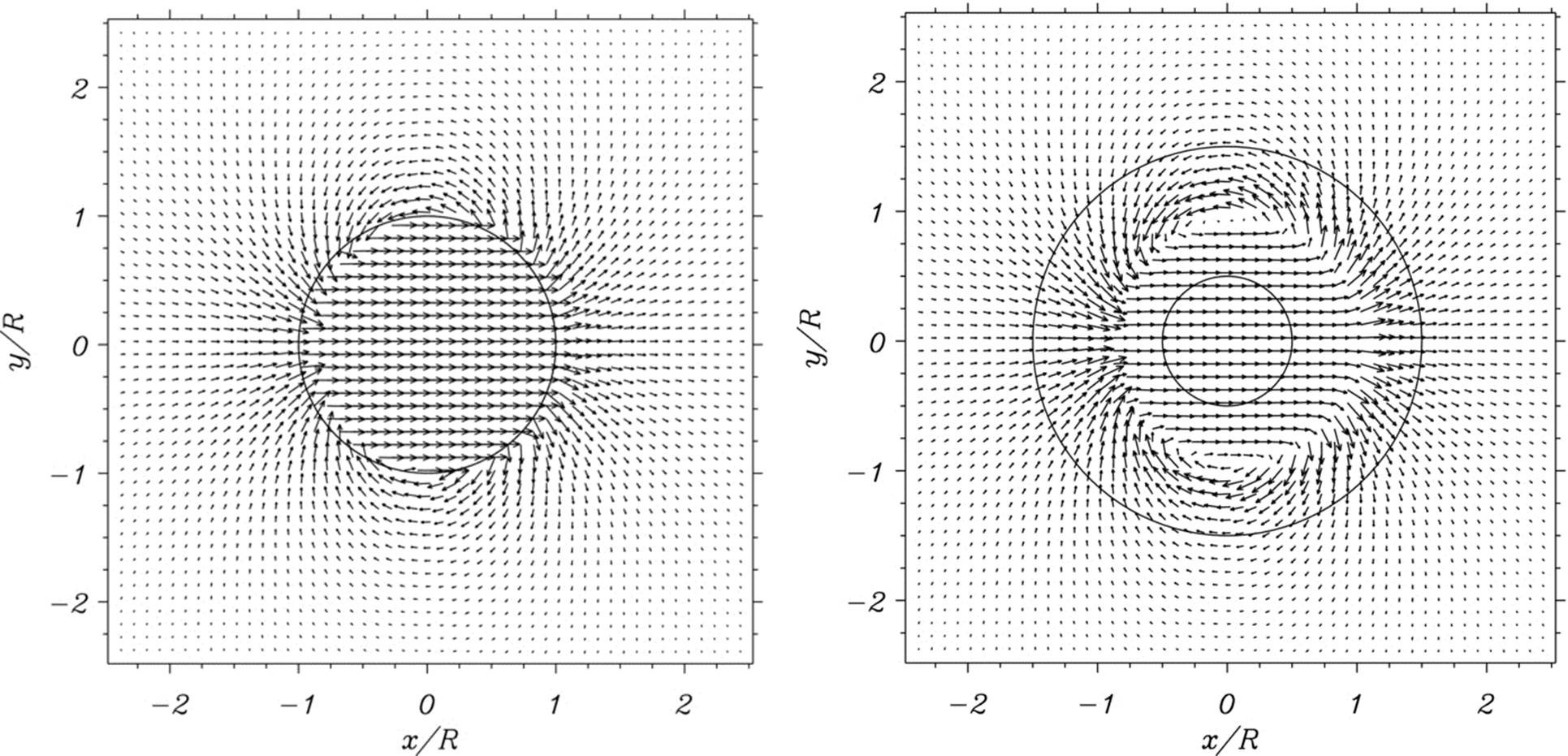}
\end{center}
\caption{The left panel shows a cross-cut of the kink wave displacement field (arrows), perpendicular to the flux tube axis. The internal and external densities are discontinuous at the boundary, indicated by the circle. The right panel shows a kink ``quasi-mode'' displacement field (arrows) in a non-uniform tube with $l/R = 1$. The non-uniform density layer between the internal and external plasma is indicated by the annulus. The resultant resonance will cause the transverse motion of the kink wave to be mode converted into localized non-axisymmteric ($m=1$) torsional Alfv\'{e}n motion within the inhomogeneous layer, i.e., rotational motion will be amplified as the displacement field evolves. Image adapted from \citet{Goo14}.}
\label{kink_rot_comb}
\end{figure*}

Since spicules are rooted in intergranular lanes where such vortex motion occurs, it is natural to assume that torsional motion should be common in all of these structures. To search for this, the high resolution capabilities of CRISP and the TRI-Port Polarimetric Echelle-Littrow (TRIPPEL) spectrograph, both based at the SST, were exploited by \citet{DeP12}. They were successful in actually resolving red-blue Doppler velocity asymmetries across the width of spicules in both H$\alpha$ and Ca~{\sc{ii}}~H data, interpreting this as the clear signature of torsional Alfv\'{e}n waves (see Figure~\ref{torsion_doppler}). Hence, a more complete picture is now emerging of spicule dynamics. These chromospheric and interface region magnetic structures support at least three distinct types of motion, i.e., field-aligned flows along with both kink and torsional waves \citep{Sek13}. Modelling all these motions as independent, \citet{DeP12} estimated that the best fit to observed data via Monte-Carlo simulations required field-aligned flows of $50-100$~km{\,}s$^{-1}$, kink velocities of $15-20$~km{\,}s$^{-1}$ and torsional motions of $25-30$~km{\,}s$^{-1}$. A drawback of this forward model is that by assuming these motions are independent, it completely neglects the physical magneto-fluid behaviour of the plasma. Realistically, these motions are coupled and this connectivity should be taken into account when interpreting data.

In the particular case of the $m=1$ kink wave, it was pointed out by \citet{Goo14} that the velocity field of this MHD wave mode, even without the presence of other modes, is actually a combination of both transverse and rotational motion. Although in the long wavelength limit the internal velocity field is purely transversal, as shown in Figure~\ref{kink_rot_comb} (left), the external field is dipolar in nature and could certainly contribute to rotational motion measured in observational data through artifacts of line-of-sight integration.
Furthermore, this rotational motion can be significantly enhanced for the kink wave through the process of resonant absorption shown in Figure~\ref{kink_rot_comb} (right). In essence, this mechanism causes the transverse energy of a kink wave to be channelled to the $m=1$ torsional Alfv\'{e}n wave in an inhomogeneous intermediate layer between the internal and external plasma where the kink wave frequency matches the local Alfv\'{e}n frequency.

Unlike the $m=0$ torsional Alfv\'{e}n wave, the $m=1$ torsional Alfv\'{e}n wave does not have azimuthal symmetry \citep[see, e.g.,][]{Spr81} and would therefore not look the same to an observer from any given line-of-sight. \citet{Goo14} pointed out that the $m=1$ rotational motions could produce very similar Doppler signatures to the $m=0$ torsional Alfv\'{e}n wave if the observer's line-of-sight is approximately perpendicular to the bulk transverse kink motion. Hence, if a spicule is observed to have a clear periodic transverse motion, indicating the presence of a kink wave, the Doppler signal across its width is likely to have a significant contribution from its $m=1$ rotational motion. Although \citet{DeP11} mainly interpreted red/blue Doppler asymmetries as being due to the $m=0$ torsional Alfv\'{e}n wave, they did not discount the possibility of $m>0$ rotational motion being present. This offers a great opportunity for both theorists and observers to understand the interplay between different MHD wave modes and flows in fine-scale chromospheric waveguides. Here, the transverse structure will be spatially resolved, with simultaneous imaging and Doppler data combined to give an accurate insight into the true nature of the plasma dynamics at work.

So in summary, MHD wave mode identification in chromospheric waveguides has caused much and often heated debate. However, without this foundation knowledge we cannot truly understand the various important aspects related to their contribution to the total energy budget of the solar atmosphere and their possible contribution to plasma heating. In this regard, the following Section~\ref{energy flux} reviews progress in determining more accurate energy flux estimations of specific MHD wave modes observed in the chromosphere.

\section{MHD wave mode energy flux}
\label{energy flux}
As discussed in Section~\ref{kink}, \citet{DeP07} interpreted the transverse motions of spicules not as MHD kink waves, but as bulk Alfv\'{e}n waves. This led them to also make energy flux estimates using the expression for such waves, i.e.,
\begin{equation}
E=\frac{1}{2}\rho v^2 c_A \ ,
\label{alfven_energy}
\end{equation}
where, $v$ is the maximum transverse velocity amplitude and the  factor $1/2$ comes from the time-averaged energy over one complete period.

Equation (\ref{alfven_energy}) is only valid under the assumption of plasma homogeneity, which results in the equipartition between kinetic (KE) and magnetic (ME) energy. Furthermore, \citet{DeP07} assumed that the energy associated with the transverse waves in spicules was of the same order elsewhere in the chromosphere and interface region, i.e., in regions where transverse waves were not observable due to the intensity S/N ratio being too low. Hence, \citet{DeP07} assumed the filling factor of the wave flux energy to be unity, even though spicules themselves are estimated to have a filling factor of no more than about 5\% in the chromosphere \citep{Mak03, Kli12}. This has led to a number of serious criticisms by various authors \citep[e.g.,][]{Erd07, Van08, Van14, Goo13}.

For the reasons discussed in Section~\ref{kink}, it was pointed out by \citet{Erd07} and \citet{Van08} that these transverse oscillations are more accurately interpreted as MHD kink waves. \citet{Van08} also went on to point out that kink wave energy flux would be strongly influenced by the filling factor of the plasma structure it was propagating through. In fact, for an overdense flux tube relative to the ambient plasma, as can be assumed for spicules, the kink wave  energy flux takes a maximum value within the tube itself and decays in the external region. The rate of decay depends on both the longitudinal wavenumber and density contrast. The larger the longitudinal wavenumber or density contrast, the faster the energy flux decays as a function of distance from the tube.

Importantly, \citet{Goo13} demonstrated that for $\rho_i \ne \rho_e$, a kink wave has no local equipartition of KE and ME. In the long wavelength approximation with $B_i=B_e$, the ratio of ME to KE inside the tube is,
\begin{equation}
\left(\frac{\mathrm{ME}}{\mathrm{KE}}\right)_i=\frac{\rho_i+\rho_e}{2 \rho_i} \ ,
\label{ratio_in}
\end{equation}
and outside it is,
\begin{equation}
\left(\frac{\mathrm{ME}}{\mathrm{KE}}\right)_e=\frac{\rho_i+\rho_e}{2 \rho_e} \ .
\label{ratio_out}
\end{equation}
Also from \citet{Goo13}, the ratio of external to internal total energy ($\mathrm{TE}=\mathrm{KE}+\mathrm{ME}$) is,
\begin{equation}
\frac{(\mathrm{TE})_e}{(\mathrm{TE})_i}=\frac{3\rho_e+\rho_i}{3\rho_i+\rho_e} \ .
\label{ratio_energy}
\end{equation}
Equations (\ref{ratio_in}) and (\ref{ratio_out}) show that local energy equipartition is only possible if $\rho_e=\rho_i$. An overdense flux tube (i.e., $\rho_i>\rho_e$) results in $(\mathrm{ME}/\mathrm{KE})_i<1$ inside and $(\mathrm{ME}/\mathrm{KE})_e>1$ outside. Hence, KE dominates inside the tube and ME dominates outside. Note also that $(\mathrm{TE})_e/(\mathrm{TE})_i<1$, implying there is more of the total energy inside the tube than outside.

To illustrate the spatial variance in the distribution of energy with a specific numerical example, we take $\rho_i/\rho_e~=~3$. This results in  $(\mathrm{ME}/\mathrm{KE})_i \approx 0.7$ and $(\mathrm{ME}/\mathrm{KE})_e =2$. The ratio of total energies gives $(\mathrm{TE})_e/(\mathrm{TE})_i=0.6$, hence, more than half the total energy is inside the tube. Denoting the flux tube radius as $R$, it can be shown that 90\% of TE ($= \mathrm{TE}_i+\mathrm{TE}_e$) is within $2R$ of the flux tube axis, and 98.5\% is within $5R$. Hence, even for a modest density ratio, $\rho_i/\rho_e=3$, this still results in a notable localised concentration of energy in the immediate neighbourhood of the flux tube.
The larger the density ratio, the more localised this energy concentration will be. Therefore, interpreting the transverse waves found in spicules as kink waves means we have to take account of the spatially varying nature of the energy flux. \citet{Van14} derived an expression for the spatially averaged kink wave energy flux in a multi-tube system based on the calculations presented by \citet{Goo13}, assuming small filling factors ($f \lesssim 0.1$) as,
\begin{equation}
E=\frac{1}{2}f(\rho_i+\rho_e)v^2 c_k \ ,
\label{kink_energy}
\end{equation}
where all the quantities are assumed to be average values taken from kink waves propagating in a multi-flux tube system. In Equation (\ref{kink_energy}) $v$ is the average kink wave maximum transverse velocity amplitude at the various locations of the dense flux tubes. Since for a kink wave the transverse velocity amplitude decays with distance from the flux tube, $v$ in Equation (\ref{kink_energy}) has a physically distinct behaviour to the maximum transverse velocity amplitude shown in Equation (\ref{alfven_energy}) for bulk Alfv\'{e}n waves. In this simpler homogeneous plasma model the Alfv\'{e}n wave has a uniform velocity amplitude in space. Taking an upper bound spicule filling factor of $f=0.05$, \citet{Van14} applied Equation (\ref{kink_energy}) to the original bulk Alfv\'{e}n wave energy flux estimates of \citet{DeP07} (derived using Equation (\ref{alfven_energy})), and found they were reduced from $4000-7000$~W{\,}m$^{-2}$ to $200-700$~W{\,}m$^{-2}$.
This highlights the very important fact that if kink waves in an overdense solar waveguide are wrongly interpreted as bulk Alfv\'{e}n waves, it can lead to a substantial overestimation of the energy flux. In the particular case of \citet{DeP07}, the overestimation is believed to be at least an order of magnitude.

Regarding filling factors of fibrils seen on-disc, \citet{Mor12a} estimated a comparable upper bound to spicules of $4-5\%$. For the energy flux estimate of kink waves in fibrils \citet{Mor12a} only considered the energy inside the flux tubes, ignoring the external contribution. In essence this is similar to what was done by \citet{Van14} in deriving Equation (\ref{kink_energy}). Interestingly, \citet{Mor12a} estimated the kink wave energy flux to be $170 \pm 110$~W{\,}m$^{-2}$, the same order as that derived for spicules with the necessary filling factor correction by \citet{Van14}. This should not be surprising since spicules and fibrils both have similar densities, filling factors and transverse wave amplitudes.

Incompressible wave energy flux in the different form of torsional Alfv\'{e}n waves above a magnetic bright point group was estimated by \citet{Jes09} to be about 240~W{\,}m$^{-2}$, assuming magnetic bright points cover at least 1.6\% of the solar surface at any one time.  This is of the same order as that estimated for filling factor corrected kink waves in fibrils and spicules. Again, this is not unexpected since \citet{Mor13} showed that photospheric vortex motion, which is the natural driver of torsional Alfv\'{e}n waves, was also found to excite abundant chromospheric kink waves in H$\alpha$ fibrils.

\citet{Mor12a} also estimated the sausage wave energy flux in fibrils to be on the order of $460 \pm 150$~W{\,}m$^{-2}$, which is almost three times more than that found for kink waves, suggesting that compressive wave energy is more abundant than its incompressible counterpart.
This has important implications for their ultimate fate, since compressive and incompressive MHD wave modes can have quite different physical damping mechanisms and rates, as discussed previously in Section~\ref{intro}.

Here we add the caveat that such energy flux estimates, filling factor arguments aside, are based on resolved wave amplitudes only. Therefore, a substantial amount of wave energy may still be unaccounted for. In the corona, it has already been suggested by many authors \citep[e.g.,][to name but a few]{Has90,Has94,Ban98}, that measured non-thermal spectral line broadening between about $20-50$~km{\,}s$^{-1}$ could have a significant contribution from sub-resolution waves. With respect to the chromosphere, the soon to be operational Daniel K. Inouye Solar Telescope (DKIST) in Maui, with its 4m diameter telescope, will offer a much improved tool to probe the smaller scale wave dynamics than is currently available (e.g., via the 1m SST and the 0.76m DST). To put this into perspective, the 2-pixel diffraction-limited spatial resolution at Ca~{\sc{ii}}~K wavelengths obtained by the DKIST will be sub-20~km, compared with the near-100~km resolution offered by the DST. By helping us to better understand the velocity fields and fine structure of chromosphere waves, it will also provide us with more accurate energy flux estimates.

\begin{figure*}
\begin{center}
\includegraphics[angle=0,width=16cm]{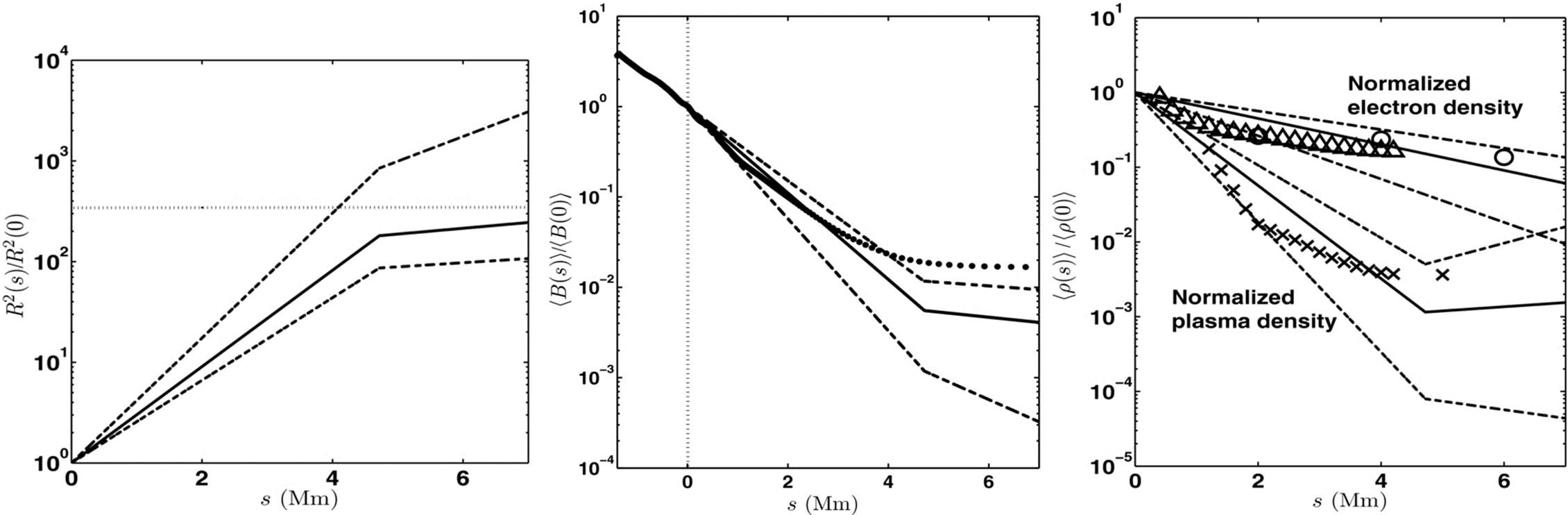}
\end{center}
\caption{The panels show magnetoseismically derived estimates of background variables along a spicule by \citet{Ver11} in comparison with results of other authors.  The left panel shows the estimated area expansion of the flux tube, normalized to unity at the spicule footpoint $s = 0$, where $s$ is the field-aligned coordinate along the spicule. The dashed lines signify the 95\% confidence bounds. The dotted horizontal line denotes the upper limit estimated by \citet{Tsu08} using Hinode/SOT data. The middle panel indicates the estimated decrease in magnetic field strength with height, normalized to unity at $s = 0$. The dashed lines also signify the 95\% confidence bounds. The dots show the average unsigned magnetic field strength from the radiative MHD simulations of \citet{DeP07} to a depth of $−1.4$~Mm. The vertical dotted line indicates the position of the photosphere. The right panel illustrates both normalized plasma and electron density. The solid line indicates the magnetoseismically determined variation in plasma density from \citet{Ver11}, with the dashed lines indicating the 95\% confidence bounds. The crosses are from Table 1 of \citet{Mak03}. Regarding the normalized electron density, the circles are from Table XIX of \citet{Bec68} and the triangles are from Table 1 of \citet{Mak03}. The estimate and uncertainties by \citet{Bjo08} are shown by the solid and dashed lines respectively. Image adapted from \citet{Ver11}.}
\label{sms_comb}
\end{figure*}

In summary, it was to be expected that from the preceding debate about chromospheric MHD wave mode identification, arguments would also arise about the actual energy flux they contain. In this Section we have highlighted some of the main differing ideas on this contentious issue. In the next Section, we go on to review how the discovery of these MHD wave modes has helped us advance the field of chromospheric magnetoseismology.

\section{Advances in chromospheric \\ magnetoseismology}
\label{CMS}
Significantly, using Hinode/SOT Ca~{\sc{ii}}~H data, \citet{He09b} measured the variation in both propagation speed and velocity amplitude of kink waves as they travelled along spicules. \citet{Ver11} exploited this detailed information for the purpose of implementing chromospheric magnetoseismology. Previously, magnetoseismology in the Sun's atmosphere was limited to TRACE observations of post-flare standing kink waves in coronal loops \citep[e.g.,][]{Nak01, Asc04}. Since it was mostly the fundamental mode that was observed in such events, this did not provide enough information for wave theorists to determine how the plasma density and magnetic field were varying along such structures. Obtaining more detailed information about field-aligned plasma inhomogeneity length scales from standing kink wave observations requires the detection of higher harmonics \citep[see, e.g.,][]{And09}, but unfortunately these were found much less frequently in the data. However, after the discovery of ubiquitous propagating kink waves in the chromosphere, this opened a whole new avenue in solar atmospheric magnetoseismology. The governing kink wave equation that had initially been derived for standing kink waves in coronal loops of longitudinally varying magnetic field and plasma density \citep[see, e.g.,][]{Rud08, And11}, could now be applied to observations of propagating waves in the lower atmosphere. The ordinary differential equation that describes the transverse velocity component of undamped kink waves in the thin tube regime is,
\begin{equation}
\frac{d^2}{d s^2}\left(\frac{v}{R}\right)+\frac{\omega^2}{c_k^2(s)}\left(\frac{v}{R}\right)=0 \ ,
\label{expand_kink}
\end{equation}
where $s$ is the magnetic field aligned co-ordinate, $\omega$ is the angular frequency, $v(s)$ is the maximum transverse velocity component, $R(s)$ is the flux tube radius and,
\begin{equation}
c^2_k(s)=\frac{B^2(s)}{\mu\left< \rho(s)\right>} \ ,
\label{kink_speed_av}
\end{equation}
where $B(s)$ is the magnetic field strength, taken to be the same inside and outside the tube, and $\left< \rho(s)\right>=[\rho_i(s)+\rho_e]/2$ is the average of the internal and external densities.

If both the maximum transverse velocity, $v(s)$, and kink speed, $c_k(s)$, are estimated from observations, then Equation (\ref{expand_kink}) can be solved for the only unknown, $R(s)$.  From the determined $R(s)$ and the thin tube magnetic flux conservation relation,
\begin{equation}
B(s)\propto \frac{1}{R^2(s)} \ ,
\label{flux_con}
\end{equation}
the variation in magnetic field, $B(s)$, along the flux tube can also be deduced. Combining $B(s)$ with the original observational estimate of $c_k(s)$, we can go back to Equation (\ref{kink_speed_av}) for determining the field aligned variation in average plasma density,
\begin{equation}
\left<\rho(s)\right>\propto \frac{B^2(s)}{c_k^2(s)} \ .
\end{equation}
Therefore, observational estimates of $v(s)$ and $c_k(s)$
allow us to determine the variation of both the magnetic
field and plasma density along solar waveguides.

\citet{Ver11} pioneered this magnetoseismological approach to find the variation of $R(s)$, $B(s)$ and $\left<\rho(s)\right>$ along a spicule (see Figure~\ref{sms_comb}). This technique was later implemented by \citet{Kur13} and \citet{Mor14a} in further investigations of mottles and spicules, respectively. In fine-scale plasma structures of near-resolution width such as spicules and mottles, $R(s)$ can be difficult to determine from intensity information alone \citep[see, e.g.,][]{DeF07}.
Also, traditional methods for determining plasma density and magnetic field strengths in the chromosphere through spectroscopy \citep[e.g.,][]{Mak03, Bjo08} and polarimetry \citep[e.g.,][]{Tru05, Cen10} have their own inherent problems. Hence, magnetoseismology provides a much needed complementary approach in determining near- (or even sub-) resolution structuring of the chromosphere.

Note that for Equation (\ref{expand_kink}) the propagating wave envelope, $v(s)$, is independent of $\omega$ since the effect of frequency-dependent damping is not included. \citet{Ver11} pointed out that if damping was present this would result in underestimating the rate of change of $R(s)$ and hence the other quantities, $B(s)$ and $\left<\rho(s)\right>$. In fact, the damping rate of kink waves and other MHD wave modes in the chromosphere are of much interest, but so far little is known. In contrast, the damping rates of post-flare/CME standing kink waves in the corona have been very well studied. In a statistical analysis of 52 standing kink wave events in coronal loops, using combined TRACE and SDO/AIA data, \citet{Ver13} found most quality factors were in the range $\tau/P \approx 1-4$, where $\tau$ is the damping time and $P$ the period. A widely supported physical mechanism to explain this is resonant absorption \citep[see the review by][]{Goo11}. In an overdense flux tube resonant behaviour is often modelled analytically by the inclusion of an annulus at the boundary where the value of density decreases continuously from $\rho_i$ to $\rho_e$. Hence, an Alfv\'{e}n continuum is introduced into the flux tube via the creation of a boundary layer.
Since the kink frequency is between that of the internal and external Alfv\'{e}n frequencies, at some position in the boundary layer the kink frequency will match that of the local Alfv\'{e}n frequency and a resonance will occur. This causes the kink wave to be mode converted to the $m=1$ torsional Alfv\'{e}n wave in the boundary layer, resulting in the observed kink transverse motion becoming damped. Analytically, this process can be described most easily in the thin tube thin boundary (TTTB) approximation, which predicts an exponential kink wave damping rate with a quality factor,
\begin{equation}
\frac{\tau}{P}=F\, \frac{R}{l}\frac{\rho_i+\rho_e}{\rho_i-\rho_e} \ ,
\label{qual_fac}
\end{equation}
where $l$ is the width of the inhomogeneous density layer, $R$ is the flux tube radius and the factor $F$ depends on the functional form chosen for the decrease in density between $\rho_i$ and $\rho_e$. To give a particular example, choosing a sinusoidal decrease results in $F=2/\pi$. It can be seen from Equation (\ref{qual_fac}) that the damping time can be reduced by increasing the boundary layer width relative to the flux tube radius (larger $l/R$), and also by increasing the internal/external density contrast (larger $\rho_i/\rho_e$). The damping rate predicted by Equation (\ref{qual_fac}) has been exploited to determine the cross-field variation in plasma density in solar atmospheric waveguides through observed damping rates. Primarily, this has been attempted for standing and propagating kink waves in the corona \citep[see, e.g.,][]{Asc03, Arr07, Ver10}. Now there is such an extensive data set for the damping rates of coronal kink waves, even more advanced statistical models are now being employed \citep[e.g.,][]{Arr13, Ver13, Arr14}. In contrast, to date there have only been a few attempts at estimating the in situ damping rates of kink waves in the chromosphere \citep[e.g.,][]{Kur12, Mor14a}.

The basis of support for the mechanism of resonant absorption to explain observed kink wave damping is mostly founded on two separate arguments. Firstly, expected order-of-magnitude values for both viscosity and resistivity in the corona would not account for the reasonably fast damping rates \citep[e.g.,][]{Nak99}. Secondly, flux tubes in the solar atmosphere are unlikely to have perfect discontinuities in Alfv\'{e}n speed at their boundaries as idealised by \citet{Edw83}. Hence, inclusion of a more realistic Alfv\'{e}n continuum between the internal and external plasma would naturally introduce a resonant layer for the kink wave. For a broadband frequency driver, resonant absorption and the resultant process of mode conversion will cause $m=1$ torsional Alfv\'{e}n waves to be excited on many magnetic surfaces which will then phase mix. This may lead to Kelvin-Helmholtz instabilities between neighbouring magnetic surfaces, which in turn will generate smaller length scales at which heating becomes more efficient \citep[e.g.,][]{Ofm98,Ter08, Ant14}.

Importantly, such broadband frequency propagating kink waves were discovered in coronal loops with the CoMP instrument by \citet{Tom07}.  This inspired theorists to model the process of resonant absorption for propagating kink waves. Initial work in this area was by \citet{Ter10}, who found that the damping length ($L_D$) for kink waves in the TTTB approximation is inversely proportional to the frequency, $f$,
\begin{equation}
L_D=\frac{1}{f}\left(\frac{\tau}{P}\right)c_k \ .
\label{damp_length}
\end{equation}
Hence, the process of resonant damping of kink waves acts like a low-pass filter in solar atmospheric waveguides.

\citet{Ver10} fitted Equation (\ref{damp_length}) to CoMP data of broadband frequency kink waves propagating in coronal loops with $c_k \approx 600$~km{\,}s$^{-1}$ and found $\tau/P \approx 2.7$, consistent with the range expected for standing kink waves ($\tau/P \approx 1-4$). Such frequency dependent damping should be detectable in velocity power spectra as a function of height in the solar atmosphere. In fact, \citet{Mor14} searched for this by comparing velocity power spectra from both the chromosphere and corona using the ROSA/DST and CoMP instruments, respectively. Interpolating over the height spanned by the interface region between the ROSA/DST and CoMP data (approximately $10-15$ Mm), it was found that the damping length of kink waves in the interface region was about 12.5\% of that previously estimated in the corona.
This provided tentative evidence of greatly enhanced damping of propagating kink waves in the Sun's lower atmosphere. \citet{Mor14} suggested that this could be caused by a combination of smaller quality factors and lower kink speeds in the interface region. The recently launched IRIS spacecraft, which is specifically designed to give the highest spatial/temporal resolution yet in interface region  spectral lines, will be an invaluable tool to push forward from these initial studies by \citet{Mor14} and actually measure the changing properties of kink waves as they traverse this fascinating region. Certainly, a whole new era of chromospheric and interface region  magnetoseismology is opening up before us.

\section{Summary}
\label{summary}
Until the launch of Hinode in 2006, solar atmospheric MHD wave observers and theorists were almost exclusively focused on the corona. However, with the discovery of ubiquitous transverse waves in chromospheric and interface region spicules by \citet{DeP07}, this branch of solar atmospheric wave research gained a whole new lease of life. Now there is vigorous debate about which particular MHD wave mode (or modes) are being observed in the Sun's lower atmosphere. Moreover, this has naturally lead to intense discussions about how best to quantify their associated energy flux, knowledge of which is crucial for understanding the contribution of waves to plasma heating.
The plethora of high spatial/temporal ground- and space-based chromospheric imaging and spectroscopic data now available has also allowed the first magnetoseismic studies of this fascinating and complex region. This has proved an especially useful complementary tool to probe the fine-scale plasma structuring of the chromosphere. With the more traditional methods of spectroscopy and polarimetry, it is very difficult to estimate even average plasma densities and magnetic field strengths in near-resolution width and short-lived chromospheric features such as spicules, fibrils and mottles, far less how they vary in time and space. Now magnetoseismology really has something to offer in this regard. In the context of all-encompassing studies of MHD wave propagation and energy deposition throughout the whole solar atmosphere, it is becoming increasingly clear that chromospheric waves play a vital role. In conclusion, a whole new era of chromospheric MHD wave research has truly unfolded, offering fantastic opportunities and challenges to both theorists and observers alike.

%
%
%
%
%
%
%

\begin{acknowledgments}
G.V. acknowledges the support of the Leverhulme Trust (UK).
D.B.J. wishes to thank the UK Science and Technology
Facilities Council (STFC) for the award of an Ernest
Rutherford Fellowship alongside a dedicated Research Grant. 
The authors also acknowledge R.J. Morton for collating the data used in Table~\ref{kink_tab}.
\end{acknowledgments}

\end{article}
%
%
%
%
%
%
%
%


\end{document}